\documentclass[aps,prl,groupedaddress,showpacs,floatfix]{revtex4}

\usepackage{amsfonts,amssymb}

\newlength{\Taille}

\newcommand{\flechebas}[1]{
  \settoheight{\unitlength}{\mbox{$#1$}}
  \settowidth{\Taille}{\mbox{~${\scriptstyle #1}$}}
  \addtolength{\unitlength}{4ex}
  \begin{picture}(0,1)
    \put(0,1){\vector(0,-1){1}}
    \put(0,0.5){\makebox(0,0){${\scriptstyle #1}$ \hspace{\the\Taille}}}
  \end{picture}}
\newcommand{\flechehaut}[1]{
  \settoheight{\unitlength}{\mbox{$#1$}}
  \settowidth{\Taille}{\mbox{~${\scriptstyle #1}$}}
  \addtolength{\unitlength}{4ex}
  \begin{picture}(0,1)
    \put(0,0){\vector(0,1){1}}
    \put(0,0.5){\makebox(0,0){\hspace{\the\Taille}${\scriptstyle #1}$ }}
  \end{picture}}
\newcommand{\flechedroite}[1]{
  \settowidth{\unitlength}{\mbox{$#1$}}
  \settoheight{\Taille}{\mbox{${\scriptstyle #1}$}}
  \addtolength{\Taille}{1ex}
  \addtolength{\unitlength}{4ex}
  \raisebox{0.5ex}{
  \begin{picture}(1,0)
    \put(0,0){\vector(1,0){1}}
    \put(0.5,0){\makebox(0,0){${\scriptstyle #1}$ \vspace{\the\Taille}}}
  \end{picture}}}
\newcommand{\flechegauche}[1]{
  \settowidth{\unitlength}{\mbox{$#1$}}
  \settoheight{\Taille}{\mbox{${\scriptstyle #1}$}}
  \addtolength{\Taille}{1ex}
  \addtolength{\unitlength}{4ex}
  \raisebox{0.5ex}{
  \begin{picture}(1,0)
    \put(1,0){\vector(-1,0){1}}
    \put(0.5,0){\makebox(0,0){${\scriptstyle #1}$ \vspace{\the\Taille}}}
  \end{picture}}}

\newtheorem{theorem}{Theorem}

\def\endprf{\hfill  {\vrule height6pt width6pt depth0pt}\medskip}
\newenvironment{proof}{\noindent {\bf Proof} }{\endprf\par}

\begin{document}

\title{{Mapping a Massless Scalar Field Theory on a Yang-Mills Theory: Classical Case}}

\author{Marco Frasca}
\email[]{marcofrasca@mclink.it}
\affiliation{Via Erasmo Gattamelata, 3 \\ 00176 Roma (Italy)}

\date{\today}

\begin{abstract}
We analyze a recent proposal to map a massless scalar field theory onto a Yang-Mills theory at classical level. It is seen that this mapping exists at a perturbative level when the expansion is a gradient expansion. In this limit the theories share the spectrum, at the leading order, that is the one of an harmonic oscillator. Gradient expansion is exploited maintaining Lorentz covariance by introducing a fifth coordinate and turning the theory to Euclidean space. These expansions give common solutions to scalar and Yang-Mills field equations that are so proved to exist by construction, confirming that the selected components of the Yang-Mills field are indeed an extremum of the corresponding action functional.
\end{abstract}


\maketitle

\section{Introduction}
\label{sec:introduction}

In a recent paper \cite{fra1} we proposed a theorem mapping classical solutions of a quartic massless scalar field theory on the solutions of a Yang-Mills theory. This result is relevant as, if we are able to realize this mapping, knowing the spectrum of the scalar theory in the infrared \cite{fra2}, we would get the spectrum of Yang-Mills theory in the same limit. The proof of this theorem was based on the functionals of the two theories. A subset of components of the Yang-Mills field was selected and taken to be all equal. The two functionals are then seen to coincide.

This proof cannot be considered complete for the evident reason that if one selects a small subset of the full set of connections, it is not granted that an extremum is recovered, questioning in this way the very existence of such mapping solutions between the two theories. This criticism would be completely sound if a set of partial differential equations could not be reduced to a set of ordinary differential equations but this is a standard technique and a starting point to make the proof complete.

Being these solutions crucial in some way to understand quantum field theory for Yang-Mills fields, we show in this paper that such solutions indeed exist and that can be safely used to build a quantum field theory in the infrared, proving that all the conclusions given in \cite{fra1} are indeed correct.

\section{Setting the Problem}
\label{sec:setting}

In ref.\cite{fra1} we stated the following theorem:

\begin{theorem}[Mapping]
\label{teo1}
An extremum of the action
\begin{equation}
    S = \int d^4x\left[\frac{1}{2}(\partial\phi)^2-\frac{\lambda}{4}\phi^4\right]
\end{equation}
is also an extremum of the SU(N) Yang-Mills Lagrangian when 
we properly choose $A_\mu^a$ with some components being zero and all others being equal, and
$\lambda=Ng^2$, being $g$ the coupling constant of the Yang-Mills field.
\end{theorem}

and the following proof was presented

\begin{proof} We show the existence of the mapping for SU(2) but the proof can be straightforwardly extended to any group given the product between its structure constants. The proof of this theorem is straightforwardly obtained by 
a proper substitution of the extremum of the scalar field action into the Yang-Mills action. For the latter we take 
\begin{eqnarray}
    S&=&\int d^4x\left[\frac{1}{2}\partial_\mu A^a_\nu\partial^\mu A^{a\nu}
    +\partial^\mu\bar c^a\partial_\mu c^a\right] \\ \nonumber
    &-&\int d^4x\left[gf^{abc}\partial_\mu A_\nu^aA^{b\mu}A^{c\nu}
	+\frac{g^2}{4}f^{abc}f^{ars}A^b_\mu A^c_\nu A^{r\mu}A^{s\nu}
	+gf^{abc}\partial_\mu\bar c^a A^{b\mu}c^c\right].
\end{eqnarray}
where we have taken a coupling with an external field (ghost) $\bar c^a, c^a$ that will turn out to be useful in the following. In our case one has
\begin{equation}
   f^{abc}f^{ars}=\epsilon_{abc}\epsilon_{ars}=\delta_{br}\delta_{cs}-\delta_{bs}\delta_{cr}
\end{equation}
being $\epsilon_{abc}$ the Levi-Civita symbol. This gives for the quartic term
\begin{equation}
   V(A)=(A_{\mu}^aA^{a\mu})^2-(A_\mu^aA^{a\nu})(A^{b\mu}A^b_\nu).
\end{equation}
Now we adopt Smilga's choice \cite{smi} taking for the components of the Yang-Mills field $A_1^1=A_2^2=A_3^3=\phi$ and all others being zero, One gets immediately
\begin{equation}
   V(A)=-6\phi^4=-V(\phi)
\end{equation}
and the mapping exists. We notice the factor 6 that compensates for the analogous factor appearing in the kinematic term and generates the proper 't~Hooft coupling. 

With the Smilga's choice, the obtained mapping annihilates also the coupling with the ghost field that in this way is shown to decouple from the Yang-Mills field. For consistency
reasons we are taking, also for the ghost field, all equal components.

Finally, we can write down the mapped action as
\begin{equation}
    S=-3\int d^4x\left[\frac{1}{2}\partial_\mu\phi\partial^\mu\phi 
    +\partial\bar c\partial c\right]
    +3\int d^4x\frac{2g^2}{4}\phi^4.
\end{equation}
\end{proof}

Above criticism is easy to be understood looking at this proof. We have selected a particular solution, $A_1^1=A_2^2=A_3^3=\phi$, but there is no way to be sure that is really an extremum for the connections. This puts a serious doubt about the very existence of such solutions. Indeed, in the proof we omitted the gauge fixing term and this can make clearer why the above proof is at least incomplete. So, let us write down the equations of motion for the Yang-Mills field
\begin{equation}
\partial^\mu\partial_\mu A^a_\nu-\left(1-\frac{1}{\alpha}\right)\partial_\nu(\partial^\mu A^a_\mu)+gf^{abc}A^{b\mu}(\partial_\mu A^c_\nu-\partial_\nu A^c_\mu)+gf^{abc}\partial^\mu(A^b_\mu A^c_\nu)+g^2f^{abc}f^{cde}A^{b\mu}A^d_\mu A^e_\nu = 0.
\end{equation}
It is easy to realize that, keeping the gauge fixing term, components that are chosen to be zero appears to have non-zero contributions into the equations producing in this way an inconsistency. This inconsistency can be removed in two ways: Fixing the gauge setting $\alpha=1$ (Lorenz gauge) or taking solutions that have no spatial dependence. Working directly with functionals, it is seen that the gauge fixing term is reduced to $\sum_{i=1}^3(\partial_iA_i^i)^2$ that would produce right equations. But here we have fallen back to above criticism.

There is a point here to be noticed. The result given in ref.\cite{fra1} does not rely strictly on a requirement of spatial dependence of these mapping solutions as all we do is a gradient expansion. In a gradient expansion we can prove that the two theories indeed map at least at the leading order. We will analyze higher order corrections to Yang-Mills equations to see how this inconsistency may arise. The conclusion will be that the mapping always holds in the limit $g\rightarrow\infty$.

Theorem's formulation makes no claim of generality but requires the existence of at least a set of common solutions between the two theories. Mapping in its fully generality can be obtained perturbatively in a $1/g$-expansion, holding for $g\rightarrow\infty$, that is what one needs in the low energy limit.

\section{Gradient Expansion for a Massless Scalar Field Theory}
\label{sec:scalar}

In order to give an understanding of the technique and to get the main equations to work with, we analyze the case a massless quartic scalar field. We will have
\begin{equation}
    S = \int d\tau d^4x\left[\frac{1}{2}(\partial\phi)^2-\frac{\lambda}{4}\phi^4\right]
\end{equation}
where we consider a time $\tau$ and a four-dimensional Euclidean manifold. Euler-Lagrange equations will yield
\begin{equation}
    \partial_\tau^2\phi-\Delta_2\phi+\lambda\phi^3=0
\end{equation}
being $\Delta_2=\sum_{i=1}^4\partial_i^2$. This equation admits a class of exact solutions given by
\begin{equation}
\label{eq:exact}
    \phi(x)=\mu\left(\frac{2}{\lambda}\right)^{1 \over 4}{\rm sn}(p\cdot x+\varphi,i)
\end{equation}
when
\begin{equation}
    p^2=\mu^2\left(\frac{\lambda}{2}\right)^{1 \over 2}
\end{equation}
being $\rm sn$ Jacobi snoidal function, $\mu$ and $\varphi$ two arbitrary constants. This is a class of massive solutions when $\mu\ne 0$ otherwise we get the trivial solution.

We introduce a gradient expansion in the following way. We rescale time as $\theta = \sqrt{\lambda}\tau$ and expand
\begin{equation}
   \phi = \sum_{n=0}^\infty\frac{1}{\lambda^n}\phi_n
\end{equation}
and this will produce the following set of equations to be solved
\begin{eqnarray}
   \partial_\theta^2\phi_0+\phi_0^3&=&0 \\ \nonumber
   \partial_\theta^2\phi_1+3\phi_0^2\phi_1&=&\Delta_2\phi_0 \\ \nonumber
   \partial_\theta^2\phi_2+3\phi_0^2\phi_2=&=&-3\phi_0\phi_1^2+\Delta_2\phi_1 \\ \nonumber
   &\vdots&.
\end{eqnarray}
So, we can use above exact solutions to solve the leading order equation as
\begin{equation}
\label{eq:phi0}
    \phi_0=\mu\left(\frac{2}{\lambda}\right)^{1 \over 4}{\rm sn}(p_0\tau+\varphi,i)
\end{equation}
being $p_0=\mu\left(\frac{\lambda}{2}\right)^{1 \over 4}$. We removed normalization using also the fact that $\mu\rightarrow \mu/\lambda^{1 \over 4}$. This is also a solution for the full set of equations as it should. For a gradient expansion, the two arbitrary constants can depend on space variables and the solution (\ref{eq:exact}) can be recovered.

From the above computation one can see that a gradient expansion is a strong coupling expansion being only defined when $\lambda\rightarrow\infty$ \cite{fra3,fra4}.

Finally, we can use the Fourier series of the solution of the leading order equation to obtain
\begin{equation}
    {\rm sn}(p_0\tau,i)=\frac{2\pi}{K(i)}\sum_{n=0}^\infty\frac{(-1)^ne^{-(n+\frac{1}{2})\pi}}{1+e^{-(2n+1)\pi}}
    \sin\left[(2n+1)\frac{\pi p_0}{2K(i)}\tau\right]
\end{equation}
that shows that we have a set of plane waves with frequencies $\omega_n=(2n+1)\frac{\pi p_0}{2K(i)}$. If we could prove that the classical limit is recovered in the strong coupling limit of the corresponding quantum field theory, this would be the spectrum of the theory. This is what we did in \cite{fra1,fra2}.

\section{Mapping Theories}
\label{sec:mapping}

The same technique of a gradient expansion can be applied to Yang-Mills equations. For the sake of simplicity we take SU(2) as a gauge group. Then we take
\begin{eqnarray}
   A^a_\nu &=& \sum_{n=0}^\infty \frac{1}{g^n}A^{a(n)}_\nu \\ \nonumber
   \theta &=& g\tau
\end{eqnarray}
and after time rescaling we can write Yang-Mills equations as
\begin{eqnarray}
\partial_\theta^2 A^a_\nu-\frac{1}{g^2}\Delta_2A^a_\nu& & \\ \nonumber
-\frac{1}{g}\left(1-\frac{1}{\alpha}\right)\partial_\nu(\partial_\theta A^a_0)
+\frac{1}{g^2}\left(1-\frac{1}{\alpha}\right)\partial_\nu(\partial_iA^a_i)& & \\ \nonumber
+f^{abc}A^b_0\partial_\theta A^c_\nu
-\frac{1}{g}f^{abc}A^b_0\partial_\nu A^c_0
+\frac{1}{g}f^{abc}A^b_i\partial_i A^c_\nu
-\frac{1}{g}f^{abc}A^b_i\partial_\nu A^c_i& & \\ \nonumber
+f^{abc}\partial_\theta(A^b_0 A^c_\nu)
-\frac{1}{g}f^{abc}\partial_i(A^b_i A^c_\nu)
+f^{abc}f^{cde}A^{b\mu}A^d_\mu A^e_\nu = 0.
\end{eqnarray}
Now we separate the component $\nu=0$ from those with $\nu=k$ with $k=1,2,3$ and we will get the following two sets of equations
\begin{eqnarray}
\partial_\theta^2 A^a_0-\frac{1}{g^2}\Delta_2A^a_0& & \\ \nonumber
-\left(1-\frac{1}{\alpha}\right)\partial_\theta^2 A^a_0
+\frac{1}{g}\left(1-\frac{1}{\alpha}\right)\partial_\theta(\partial_iA^a_i)& & \\ \nonumber
+\frac{1}{g}f^{abc}A^b_i\partial_i A^c_0
-f^{abc}A^b_i\partial_\theta A^c_i& & \\ \nonumber
+f^{abc}\partial_\theta(A^b_0 A^c_0)
-\frac{1}{g}f^{abc}\partial_i(A^b_i A^c_0)
+f^{abc}f^{cde}A^{b\mu}A^d_\mu A^e_0 = 0
\end{eqnarray}
and
\begin{eqnarray}
\partial_\theta^2 A^a_k-\frac{1}{g^2}\Delta_2A^a_k& & \\ \nonumber
-\frac{1}{g}\left(1-\frac{1}{\alpha}\right)\partial_k(\partial_\theta A^a_0)
+\frac{1}{g^2}\left(1-\frac{1}{\alpha}\right)\partial_k(\partial_iA^a_i)& & \\ \nonumber
+f^{abc}A^b_0\partial_\theta A^c_k
-\frac{1}{g}f^{abc}A^b_0\partial_k A^c_0
+\frac{1}{g}f^{abc}A^b_i\partial_i A^c_k
-\frac{1}{g}f^{abc}A^b_i\partial_k A^c_i& & \\ \nonumber
+f^{abc}\partial_\theta(A^b_0 A^c_k)
-\frac{1}{g}f^{abc}\partial_i(A^b_i A^c_k)
+f^{abc}f^{cde}A^{b\mu}A^d_\mu A^e_k = 0
\end{eqnarray}
from which we can immediately read out the perturbation equations at all orders. At leading order we will get
\begin{equation}
\frac{1}{\alpha}\partial_\theta^2 A^{a(0)}_0
-f^{abc}A^{b(0)}_i\partial_\theta A^{c(0)}_i+f^{abc}\partial_\theta(A^{b(0)}_0 A^{c(0)}_0)
+f^{abc}f^{cde}A^{b\mu (0)}A^{d(0)}_\mu A^{e(0)}_0 = 0
\end{equation}
and
\begin{equation}
\partial_\theta^2 A^{a(0)}_k
+f^{abc}A^{b(0)}_0\partial_\theta A^{c(0)}_k
+f^{abc}\partial_\theta(A^{b(0)}_0 A^{c(0)}_k)
+f^{abc}f^{cde}A^{b\mu(0)}A^{d(0)}_\mu A^{e(0)}_k = 0.
\end{equation}
Now we see that at the leading order, for SU(2), if adopt a Smilga's choice \cite{fra1,smi}, $A_1^1=A_2^2=A_3^3=\phi_0(t)$, where $\phi_0(t)$ is given by eq.(\ref{eq:phi0}) and $\lambda=2g^2$, and all other components being zero, we have solved our leading order equations proving that mapping indeed exists at least till order $O(1/g)$. Of course, there is no limitation due to the gauge group and this result is always true. {\sl These solutions are exact solutions of the Yang-Mills equations when we do not ask for any spatial dependence}. With these solutions, the mapping theorem given above is already true. But it is also true perturbatively when the coupling is taken infinitely large. This is already enough to fully support all the conclusions of ref.\cite{fra1}. 

We want to go further and prove that the mapping theorem is perturbatively exact when spatial dependence is retained. In order to see this, we check the next to leading order equations. These are given by
\begin{eqnarray}
\partial_\theta^2 A^{a(1)}_0
-\left(1-\frac{1}{\alpha}\right)\partial_\theta^2 A^{a(1)}_0
+\left(1-\frac{1}{\alpha}\right)\partial_\theta(\partial_iA^{a(0)}_i)& & \\ \nonumber
+f^{abc}A^{b(0)}_i\partial_i A^{c(0)}_0
-f^{abc}A^{b(0)}_i\partial_\theta A^{c(1)}_i
-f^{abc}A^{b(1)}_i\partial_\theta A^{c(0)}_i& & \\ \nonumber
+f^{abc}\partial_\theta(A^{b(0)}_0 A^{c(1)}_0)
+f^{abc}\partial_\theta(A^{b(1)}_0 A^{c(0)}_0)
-f^{abc}\partial_i(A^{b(0)}_i A^{c(0)}_0)& & \\ \nonumber
+f^{abc}f^{cde}A^{b\mu(1)}A^{d(0)}_\mu A^{e(0)}_0 
+f^{abc}f^{cde}A^{{b\mu(0)}}A^{d(1)}_\mu A^{e(0)}_0 
+f^{abc}f^{cde}A^{{b\mu(0)}}A^{d(0)}_\mu A^{e(1)}_0 = 0
\end{eqnarray}
and
\begin{eqnarray}
\partial_\theta^2 A^{a(1)}_k
-\left(1-\frac{1}{\alpha}\right)\partial_k(\partial_\theta A^{a(0)}_0)& & \\ \nonumber
+f^{abc}A^{b(1)}_0\partial_\theta A^{c(0)}_k
+f^{abc}A^{b(0)}_0\partial_\theta A^{c(1)}_k \\ \nonumber
-f^{abc}A^{b(0)}_0\partial_k A^{c(0)}_0
+f^{abc}A^{b(0)}_i\partial_i A^{c(0)}_k
-f^{abc}A^{b(0)}_i\partial_k A^{c(0)}_i& & \\ \nonumber
+f^{abc}\partial_\theta(A^{b(1)}_0 A^{c(0)}_k)
+f^{abc}\partial_\theta(A^{b(0)}_0 A^{c(1)}_k)
-f^{abc}\partial_i(A^{b(0)}_i A^{c(0)}_k) \\ \nonumber
+f^{abc}f^{cde}A^{b\mu(1)}A^{d(0)}_\mu A^{e(0)}_k 
+f^{abc}f^{cde}A^{{b\mu(0)}}A^{d(1)}_\mu A^{e(0)}_k 
+f^{abc}f^{cde}A^{{b\mu(0)}}A^{d(0)}_\mu A^{e(1)}_k = 0.
\end{eqnarray}
From Smilga's choice for SU(2), being $A_0^{a(0)}=0$, we can further simplify these equations to give
\begin{eqnarray}
\alpha^{-1}\partial_\theta^2 A^{a(1)}_0
+\left(1-\frac{1}{\alpha}\right)\partial_\theta(\partial_iA^{a(0)}_i)& & \\ \nonumber
-f^{abc}A^{b(0)}_i\partial_\theta A^{c(1)}_i
-f^{abc}A^{b(1)}_i\partial_\theta A^{c(0)}_i& & \\ \nonumber
+6\phi_0^2A^{a(1)}_0 = 0
\end{eqnarray}
and
\begin{eqnarray}
\partial_\theta^2 A^{a(1)}_k+f^{abc}A^{b(1)}_0\partial_\theta A^{c(0)}_k\\ \nonumber
+f^{abc}A^{b(0)}_i\partial_i A^{c(0)}_k
-f^{abc}A^{b(0)}_i\partial_k A^{c(0)}_i& & \\ \nonumber
+f^{abc}\partial_\theta(A^{b(1)}_0 A^{c(0)}_k)
-f^{abc}\partial_i(A^{b(0)}_i A^{c(0)}_k) \\ \nonumber
+6\phi_0^2 A^{a(1)}_k = 0.
\end{eqnarray}
For these equations to hold we now assume that the parameters of the scalar field should depend on spatial variables. So, we can rewrite above equations as
\begin{eqnarray}
\alpha^{-1}\partial_\theta^2 A^{a(1)}_0
+6\phi_0^2A^{a(1)}_0 =-\left(1-\frac{1}{\alpha}\right)\partial_\theta(\delta_{ai}\partial_i\phi_0)
\end{eqnarray}
and
\begin{eqnarray}
\partial_\theta^2 A^{a(1)}_k+6\phi_0^2 A^{a(1)}_k = -f^{abc}A^{b(1)}_0\delta_{ck}\partial_\theta\phi_0.
\end{eqnarray}
These equations represent the main result of the paper as give the first order correction to a classical solution of the Yang-Mills equations in a $1/g$-expansion. From this equations we see immediately the effect of the gauge term with $\alpha$. In a Lorenz gauge we see that the two theories are fully mapped also with respect to spatial dependence. In all other cases, we recover gauge contributions at order $O(1/g)$. At higher orders we expect contributions from the Laplacian term. This means that the following is generally true
\begin{equation}
   A_\mu^a(x)=\eta_\mu^a\phi(x)+O(1/g)
\end{equation}
being $\eta_\mu^a$ a constant.


So, we have proved that the mapping theorem holds for a class of exact solutions of Yang-Mills theory selected through a proper choice of the components of the field (Smilga's choice \cite{smi}). Then, noting that Yang-Mills theory is Lorentz invariant, we can always operate a Lorentz transformation taking $\phi_0(t)$ to the exact solution (\ref{eq:exact}) confirming the proof of the mapping theorem given in \cite{fra1}. But, an essential conclusion is that the mapping holds, in a more general sense, perturbatively in a $1/g$-expansion, making fully consistent the results presented in \cite{fra1}. This is what we aimed to prove.



A by-product of our proof is that Smilga's choice selects an extremum of the Yang-Mills functional and, in a more general sense, an extremum in a $1/g$-expansion and the theories can be mapped in the limit $g\rightarrow\infty$ that is what one needs in the low-energy limit.

We note that this mapping, in quantum field theory, can only hold in the infrared limit. In the ultraviolet, due to quantum fluctuations, the correspondence is spoiled (asymptotic freedom). Anyhow, this is enough to obtain the spectrum of the theory in the required limit.

\section{Conclusions}
\label{sec:conclusions}

We have shown, by construction, that Smilga's choice select a set of solutions of Yang-Mills equations that are an extremum for the action functional and that the mapping holds in a more general sense when $g\rightarrow\infty$. This completes the proof of the mapping theorem given in \cite{fra1}. This implies that a general technique used in physics to solve partial differential equations can be straightforwardly used for Yang-Mills theory. Indeed, scalar field theory and Yang-Mills theory can be mapped each other, in fully generality, in the limit $g\rightarrow\infty$ that is what one needs working in a strong coupling regime.

\begin{acknowledgments}
I would like to thank Terence Tao for his remarks to the previous proof of the mapping theorem.
\end{acknowledgments}

\end{document}